\documentclass[prb,aps,twocolumn,superscriptaddress,floatfix,showpacs]{revtex4}
\usepackage{graphicx}
\usepackage{amssymb}
\usepackage{amsmath}
\usepackage{xspace} 
\usepackage{color}

\begin{document}

\title{Temperature effects in excitonic condensation driven by the lattice distortion}
\author{Thi-Hong-Hai Do}
\affiliation{Hanoi University of mining and geology, Duc Thang, Hanoi, Vietnam}
\author{Huu-Nha Nguyen}
\affiliation{Department of Physics, HCMC University of Science, 227 Nguyen Van Cu, Ho Chi Minh city, Vietnam}
\author{Thi-Giang Nguyen}
\affiliation{Institute of Physics, Vietnam Academy of Science and Technology, 10 Dao Tan, Hanoi, Vietnam}
\author{Van-Nham Phan}
\affiliation{Institute of Research and Development, Duy Tan University, K7/25 Quang Trung, Danang, Vietnam}

\begin{abstract}
The stability of the excitonic condensation at low temperature driven by a coupling of electrons to vibrational degrees of freedom in semimetal two-dimensional electronic system is discussed. In the framework of the unrestricted Hartree-Fock approximation, we derive a set of equations to determine both the excitonic condensate order parameter and lattice displacement self-consistently. By lowering temperature we find out a semimetal-insulator transition in the system if the coupling is large enough. The insulating state typifies an excitonic condensation accompanied by a finite lattice distortion. Increasing temperature, both excitonic condensate order parameter and the lattice distortion decrease and then disappear in the same manner. Microscopic analysis in momentum space strongly specifies that the excitonic condensate driven by the lattice distortion favours the BCS type.
\end{abstract}
\date{\today}
\maketitle

\section{Introduction}
Electron-hole pairing or excitonic condensation recently has become one of the most attractive subjects in many-particle physics~\cite{SK14,CS15}. At sufficiently low temperature, high density excitons may condense and the system stabilizes in an insulating state with respect to the spontaneous formation of a new macroscopic phase-coherent quantum state called an excitonic insulator (EI). Investigating the EI state has been intensively focused in the literature but mainly on the purely electronic manners~\cite{Ba02b,Fa08,PBF10,ZIBF12}. In doing so the coupling of electrons or excitons to the phonon was completely neglected. 

However, recently, several experiments have opened an issue that the lattice distortion is non-negligible to anticipate the EI phase in the quasi-two dimensional transition metal dichalgogenide 1$T$-TiSe$_2$. In this material, the charge density wave (CDW) has been manifested to accompany with the weak periodic lattice distortion~\cite{SMW76}. At low temperature, photoemission signatures indicate that the exciton condensate strongly associates to the CDW state~\cite{MSGMDBACMBBT10}. By measuring the thermodynamic properties of TmSe$_{0.45}$Te$_{0.55}$, Wachter and co-workers proposed a strong phonon-exciton couple in the excitonic condensate phase, forming exciton-polaron quasiparticles~\cite{Wa01,WBM04}. In their studies, the heat conductivity shows a divergence for $T\rightarrow 0$. This anomaly is analogous to the observation for $^4$He II, a typical superfluidity below $2.2$K where the heat is being carried by phonons~\cite{TT90}. In TmSe$_{0.45}$Te$_{0.55}$, at an appropriate pressure, an excitonic bound state of a $4f$ hole at the $\Gamma$-point and a $5d$ electron at the $X$-point can be created. An interaction of electron/hole with a phonon thus needs to assist that $\Gamma-X$ transfer.~\cite{Wa01} At sufficiently low temperature, those excitons condense to a superfluid forming the EI state~\cite{Wa01,WBM04}. Without any doubt, lattice distortion or phonon effects are extremely important in this kind of material, particularly, in studying the EI state. The electron/hole-phonon interaction therefore seems to be non-negligible and needs to be considered thoroughly. 

On the theoretical side, the lattice distortion causing the EI state has been studied intensively however only for the ground state, i.e., at zero temperature~\cite{KTKO13,ZFBMB13,PBF13}. In general, as a kind of superfluidity, the EI state possibly occurs at finite temperature. At high temperature it might be deformed by thermal fluctuations. Studying the influence of temperature on the EI state therefore is an important effort. {The} phase diagram specifying the EI region in the temperature-pressure plane has been experimentally {measured}, which emphasizes that the excitons condense only at low temperature with intermediate pressure around the semimetal-semiconductor transition~\cite{NW90,WBM04}. Without the electron/hole-phonon coupling, {the} temperature-pressure phase diagram of the EI state has been considered theoretically in the extended Falicov-Kimball model~\cite{IPBBF08,ZFB10,ZIBF10}. 

By considering the tight-binding formalism, the temperature effects in excitonic condensate exerting a force on the lattice generating periodic ionic displacements in 1$T$-TiSe$_2$ have been discussed~\cite{MBCAB11}. Of course, here, the mean-field form of the EI order parameter depending on temperature has been assumed from the beginning. Moreover, {the} photoemission temperature dependence measured for $1T$-TiSe$_2$ has shown that it fits quite well with the mean-field form for temperatures below the EI transition temperature~\cite{MSGMDBACMBBT10}. In the present work, we {intend to} develop the unrestricted Hartree-Fock approximation, a kind of the mean-field approach but allowing for decoupling with respect to the excitonic order parameter, to a two-dimensional two-band $f$-$c$ electron model with a coupling to the phonon degrees of freedom. This coupling results {in} a `hybridization' between $f$ and $c$ electrons. In the one-dimensional case, this model has been {studied. It shows} that at zero temperature the EI exists with respect to a lattice displacement happening only if the electron-phonon coupling is larger than a critical value~\cite{PBF13}. 

The paper is organized as follows. In Sec.~2 we introduce the two-band $f$-$c$ electronic model with a coupling to the phonon. The theoretical approach is outlined in Sec.~3, where the unrestricted Hartree-Fock approximation has been developed to the specific model mentioned in Sec.~2. Sec.~4 presents and discusses in detail numerical results. Our main conclusions can be found in Sec.~5.

\section{Two-band electron-phonon interaction model}

{The system with two spinless electronic bands involving electron-phonon coupling is modeled} by the following Hamiltonian
\begin{eqnarray}
\label{1}
\mathcal{H}&=&\sum_{\mathbf{k}}{\varepsilon}^f_{\mathbf{k}}f^\dagger_{\mathbf{k}}f^{}_{\mathbf{k}}
+\sum_{\mathbf{k}}{\varepsilon}^c_{\mathbf{k}}c^\dagger_{\mathbf{k}}c^{}_{\mathbf{k}}
+\omega_{0}\sum_{\mathbf{q}}b^\dagger_{\mathbf{q}}b^{}_{\mathbf{q}}\nonumber\\
&&+\frac{g}{\sqrt{N}}\sum_{\mathbf{kq}}\left[c^\dagger_{\mathbf{k+q}}f^{}_{\mathbf{k}}
( b^{\dagger}_{\mathbf{-q}}+ b_{\mathbf q} ) + \textrm{H.c} \right] \,,
\end{eqnarray}
where $c^\dagger_{\mathbf{k}}$ ($c^{}_{\mathbf{k}}$) and  $f^\dagger_{\mathbf{k}}$ ($f^{}_{\mathbf{k}}$) {correspond to} creation (annihilation) operators of $c$ and $f$ spinless electrons carrying momentum $\bf k$. The kinetic energy of a phonon with dispersionless energy $\omega_0$ is given in the third term in Eq.~(\ref{1}) where $b^\dagger_{\mathbf{q}}$ and $b^{}_{\mathbf{q}}$ are phonon creation and annihilation operators at momentum ${\bf q}$. Here, the electronic excitation energies are given by
\begin{equation}
\label{2}
{\varepsilon}^{f,c}_{\mathbf{k}}=\varepsilon_{}^{f,c}
-t^{f,c}_{}\gamma_{\mathbf k}-\mu \, ,
\end{equation}
where $\varepsilon^{f(c)}$ represents the local part of the $f$ ($c$) electron excitation, the   
next term $-t^{f,c} \gamma_{\mathbf k}$, with 
\begin{equation}
\gamma_{\mathbf k}=2(\cos k_x+\cos k_y),
\end{equation}
accounts for the nearest-neighbor hopping in a 2D lattice, and $\mu$ is the chemical potential. The last term in Eq.~(\ref{1}) addresses a local electron-phonon interaction (with coupling constant $g$), written in ${\bf k}$-space. 

The Hamiltonian given in Eq.~(\ref{1}) is identical to the Holstein model in which by the canonical transformations, the effective electron-hole attraction can be performed~\cite{Fr52}. {Electron-hole bound states or excitons might therefore exist} due to the coupling with the lattice - a CDW state. Like Cooper pairs in superconductors, at low temperature these excitons would prefer to form a superfluid state~\cite{CS15}. 

To mimic the situation {of TmSe$_{0.45}$Te$_{0.55}$, where the quasilocalized $4f$ state has its maximum at the $\Gamma$-point and the strongly dispersive $5d$ state has its minimum at the $X$ point, we choose $t^f<0$ and $|t^f|<1$, whereas $t^c=1$ is chosen as the unit of energy. It characterizes an indirect $c-f$ coupling assisting by the $\Gamma-X$ transfer phonon. At sufficiently low temperature, the bound pairs with finite momentum $\bf Q$ might condense, indicated by a nonzero value of the order parameter, i.e.,}
\begin{equation}
\label{3i}
d_{\bf k}= \langle c^\dag_{\bf k + \bf Q} f_{\bf k}^{} \rangle \neq 0 \, ,
\end{equation}
where $\bf Q= (\pi,\pi)$ in two dimensions. {The bound pair or exciton} in this case is considered {to have} nonzero center-mass momentum which is distance between $\Gamma$ and $X$ points. That bound pairs  raise the CDW state accompanied by a $\Gamma-X$ phonon \cite{WB13}. In the following, we {also} consider
\begin{equation}
d=\frac{1}{N}\sum_{\bf k}(\langle c^\dag_{\bf k + \bf Q} f_{\bf k}^{} \rangle+\langle f_{\bf k}^{\dag} c^{}_{\bf k + \bf Q}\rangle ),
\end{equation}
as the excitonic condensation order parameter.
\section{Unrestricted Hartree-Fock approximation}
\label{S:III}

{As a kind of Hartree-Fock approximation, the unrestricted Hartree-Fock approximation allows} decoupling with respect to the off-diagonal expectation values~\cite{SC08}, such as the excitonic order parameter, $d_{\bf k}$, in our case. In this respect, we introduce the
fluctuation operator $\delta \mathcal A= \mathcal A - \langle \mathcal A\rangle$ {for an arbitrary operator $\mathcal A$, and write the electron-phonon interaction operator in Eq.~\eqref{1} as}
\begin{eqnarray}
\label{5}
&&c^\dagger_{\mathbf{k+q}}f^{}_{\mathbf{k}} (b^\dagger_{\mathbf{-q}}+b_{\bf q})=\nonumber\\
&&+\delta(c^\dagger_{\mathbf{k+q}}f^{}_{\mathbf{k}})\delta(b^\dagger_{\mathbf{-q}} +  b_{\bf q}) -\langle c^\dagger_{\mathbf{k+q}}f^{}_{\mathbf{k}}\rangle \,  \langle b^\dagger_{\mathbf{-q}} +  b_{\bf q}\rangle
\\
&&+ \big[\langle c^\dagger_{\mathbf{k+q}}f^{}_{\mathbf{k}}\rangle (b^\dagger_{\mathbf{-q}} +  b_{\bf q})
+c^\dagger_{\mathbf{k+q}}f^{}_{\mathbf{k}} \, \langle b^\dagger_{\mathbf{-q}} +  b_{\bf q} \rangle \big]
\delta_{\bf q, \bf Q}.\nonumber
\end{eqnarray}

{Assuming that the fluctuations are small, the first term on the right hand side in Eq.~\eqref{5} can be eliminated. In this case, the Hamiltonian in Eq.~(\ref{1}) reduces to the so-called unrestricted Hartree-Fock Hamiltonian which can be separated into two parts}
\begin{equation}\label{7}
\mathcal{H}_\textrm{UHF}=\mathcal{H}_e+\mathcal{H}_{ph},  
\end{equation}
where the electronic part reads
\begin{equation}
\label{7a}
\mathcal{H}_e=\sum_{\mathbf{k}}{\varepsilon}^f_{\mathbf{k}}f^\dagger_{\mathbf{k}}f^{}_{\mathbf{k}}
+\sum_{\mathbf{k}} {\varepsilon}^c_{\mathbf{k}}c^\dagger_{\mathbf{k}}c^{}_{\mathbf{k}}+V\sum_{\bf k} \big( c^\dag_{\bf k +\bf Q} f_{\bf k} + \textrm{H.c.} \big),
\end{equation}
and
\begin{equation}\label{7b}
\mathcal{H}_{ph}=\omega_{0}\sum_{\mathbf{q}}b^\dagger_{\mathbf{q}}b^{}_{\mathbf{q}}+ \sqrt N h \big( b_{-\bf Q}^\dag + b_{- \bf Q} \big),
\end{equation}
is {the} phononic one. {Note here that the additional constant has been neglected in Eq.~\eqref{7}.} In Eq.~(\ref{7a}) $V$ reads
\begin{equation}\label{8a}
V= \frac{g}{\sqrt N} \langle b_{-\bf Q} + b^\dag_{-\bf Q} \rangle \,,
\end{equation}
indicating the hybridization between the $c$ and $f$ electrons, {on one hand it }expresses the effective bound state of electron-hole pairs, on the other hand it indicates the CDW instability like in the Holstein model in the frozen phonon approximation. The factor $h$ in Eq.~(\ref{7b}) is given by
\begin{equation}\label{8b}
h  = \frac{g}{N} \sum_{\bf k} \langle c^\dagger_{\mathbf{k+Q}}f^{}_{\mathbf{k}} + f_{\mathbf{k}}^\dag c_{\mathbf{k+Q}} \rangle \, .
\end{equation}
From {the} expressions in Eqs.~(\ref{8a}-\ref{8b}) and the Hamiltonians in Eqs.~(\ref{7a}) and (\ref{7b}) we can easily realized that both $h$ and $V$ are mutually dependent. Moreover, since $b_{\bf Q} = b_{-\bf Q}$, the field contribution $\sqrt N h \big( b_{-\bf Q}^\dag + b_{ \bf Q} \big)$ has been replaced by $\sqrt N h \big( b_{-\bf Q}^\dag + b_{ -\bf Q} \big)$. Therefore a finite lattice displacement approximating to $\langle b_{-\bf Q}^\dag + b_{ -\bf Q}\rangle$ would give rise to the formation of a CDW state connected to a doubling of the lattice unit cell. 

In order to diagonalize the {unrestricted Hartree-Fock} Hamiltonian written in Eq.~(\ref{7}), firstly, we define a new phonon operator 
\begin{equation}
\label{npo}
B_{\bf q}^\dag = b_{\bf q}^\dag + \sqrt N ({h}/{\omega_0}) \delta_{\bf q, \bf Q}\,,
\end{equation} 
to diagonalize the phononic part in Eq.~(\ref{7b}). {Meanwhile}, the electronic part can be diagonalized itself by using a Bogoliubov {transformation, where the new quasi-particle fermionic operators read} 
\begin{align}
{C}^\dagger_{1, \mathbf{k}}=&\xi^{}_{\mathbf{k}}c^\dagger_{\mathbf{k+Q}}+ \eta^{}_{\mathbf{k}}f^\dagger_{\mathbf{k}}\, , \label{9e1}\\
{C}^\dagger_{2, \mathbf{k}}=&-\eta^{}_{\mathbf{k}}c^\dagger_{\mathbf{k+Q}}+\xi^{}_{\mathbf{k}}f^\dagger_{\mathbf{k}} \,. \label{9e2}
\end{align}
{Here, the} prefactors $\xi^{}_{\mathbf{k}}$ and $\eta^{}_{\mathbf{k}}$ are chosen to satisfy $\xi^{2}_{\mathbf{k}}+\eta^{2}_{\mathbf{k}}=1$. Then finally, {we are led to} a completely diagonalized Hamiltonian
\begin{equation}
\label{9c}
{\mathcal{H}}_\textrm{eff}=\sum_{\mathbf{k}}E^{1}_{\mathbf{k}}{C}^{\dagger}_{1, \mathbf{k}}{C}_{1, \mathbf{k}}
+\sum_{\mathbf{k}}E^{2}_{\mathbf{k}}{C}^{\dagger}_{2, \mathbf{k}}{C}_{2, \mathbf{k}} +{\omega}_{0} \sum_{\bf q}  {B}^\dag_{\bf q} {B}_{\bf q} \, ,
\end{equation}
where the electronic quasiparticle energies read
\begin{equation}
\label{9d}
E^{1,2}_{\mathbf k}=\frac{{\varepsilon}^c_\mathbf{k+Q}+{\varepsilon}^f_{\mathbf k}}{2}
\mp
\frac{\textrm{sgn}({\varepsilon}^f_{\mathbf k}-{\varepsilon}^c_\mathbf{k+Q})}
{2}W_{\mathbf k}\,,
\end{equation}
and the prefactors addressed in Eqs.~(\ref{9e1}-\ref{9e2}) are given by 
\begin{align}
\label{9f}
\xi^{2}_{\mathbf{k}}=&\frac{1}{2}\left[1+\textrm{sgn}({\varepsilon}^f_{\mathbf k}-{\varepsilon}^c_\mathbf{k+Q})
\frac{{\varepsilon}^f_{\mathbf k}-{\varepsilon}^c_\mathbf{k+Q}}{W_{\mathbf{k}}}\right]\,, \\
\eta^{2}_{\mathbf{k}}=&\frac{1}{2}\left[1-\textrm{sgn}({\varepsilon}^f_{\mathbf k}-{\varepsilon}^c_\mathbf{k+Q})
\frac{{\varepsilon}^f_{\mathbf k}-{\varepsilon}^c_\mathbf{k+Q}}{W_{\mathbf{k}}}\right] \, ,
\label{9fa}
\end{align}
with
\begin{equation}
\label{9g}
W_{\mathbf k}=\sqrt{({\varepsilon}^c_{\mathbf{k+Q}}-{\varepsilon}^f_{\mathbf k})^2+4|V|^2}\,.
\end{equation}
The quadratic form of Eq.~(\ref{9c}) allows to compute all expectation values formed with 
${\mathcal H}_\textrm{dia}$, {resulting in}
\begin{eqnarray}
\label{A39}
&&\langle n^c_{\mathbf{k+Q}}\rangle=\langle c^\dagger_\mathbf{k+Q}c_\mathbf{k+Q}\rangle = 
\xi^2_{\mathbf k}f^F(E^1_{\mathbf k})+\eta^2_{\mathbf k}f^F(E^2_{\mathbf k})\,,\\[0.2cm]
\label{A40}
&&\langle n^f_{\mathbf{k}}\rangle=\langle f^\dagger_{\mathbf k}f_\mathbf{k}\rangle=
\eta^2_{\mathbf k}f^F(E^1_{\mathbf k})+\xi^2_{\mathbf k}f^F(E^2_{\mathbf k}) \,, \\[0.2cm]
&&d_{\mathbf{k}}=-[f^F(E^1_{\mathbf k})-f^F(E^2_{\mathbf k})]\textrm{sgn}(\varepsilon^f_{\mathbf k}
-\varepsilon^c_\mathbf{k+Q})\frac{V}{W_{\mathbf k}}.\label{A41} 
\end{eqnarray}
Here $f^F(E_{\bf k})=1/[1+\exp(\beta E_{\bf k})]$ is {the Fermi-Dirac distribution} function and $\beta=1/T$ is {the} inverse of the temperature.

{To calculate $ \langle b^\dag_{-\bf Q} \rangle$ in Eq.~\eqref{8a} we note here that $\langle B^\dag_{\bf q} \rangle=0$ and from Eq.~(\ref{npo}) one obtains}
\begin{equation}
\label{A46}
 \langle b^\dag_{\bf q} \rangle= \langle B^\dag_{\bf q} \rangle
 - \frac{\sqrt N  \, h}{ \omega_0}\delta_{\bf q, \bf Q} =  
 - \frac{\sqrt N \, h}{\omega_0}  \delta_{\bf q, \bf Q} \,.
\end{equation} 
{Then finally we obtain the amplitude of the} lattice displacement in the EI state for a single center-mass momentum $\mathbf{Q}$ 
\begin{equation}
\label{A48}
x^{}_\mathbf{Q}=\frac{1}{\sqrt{2N\omega_0}}
\langle b^{\dagger}_\mathbf{-Q}+ b^{}_\mathbf{Q}\rangle=-\frac{h}{\omega_0}\sqrt{\frac{2}{\omega_0}},
\end{equation}

Let us also consider the electronic one-particle spectral functions. For the $c$-electron, its spectral function $A^c_{\bf k}(\omega)$ can be found following the definition
\begin{equation}
 A^c_{\bf k}(\omega) = \frac{1}{2\pi} \int_{-\infty}^\infty \langle [c_{\bf k \sigma}(t), c^\dag_{\bf k \sigma}]_+\rangle
 e^{i\omega t} dt \, , 
\end{equation}
From the diagonal form of the effective Hamiltonian in Eq.~(\ref{9c}) one {derives}
\begin{equation}\label{A49}
A^c_{\mathbf k}(\omega)=\xi^2_{\mathbf{k-Q}}\delta(\omega-E^1_{\mathbf{k-Q}})+\eta^2_{\mathbf{k-Q}}\delta(\omega-E^2_{\mathbf{k-Q}}).
\end{equation}
In the same way, we can construct the spectral function of the $f$-electron, that reads
\begin{equation}
\label{A50}
A^f_{{\mathbf k}}(\omega)=\eta^2_{\mathbf k}\delta(\omega-E^1_{\mathbf k})+\xi^2_{\mathbf k}\delta(\omega-E^2_{\mathbf k}).
\end{equation}
Note that according to the spectral function we can evaluate the corresponding density of state by taking a summation over all {momenta} in the first Brillouin zone.


\section{Numerical results}
In this section, we present numerical results to discuss the influence of temperature on {the} excitonic condensate of the two-band electrons involving the electron-phonon interaction. For the two dimensional system consisting of $N=200\times 200$ lattice sites, the numerical results are obtained by solving self-consistently Eqs.~(\ref{8a}), (\ref{8b}), (\ref{A41}), and (\ref{A46}) starting from some guess values for $\langle b^\dag_{\bf Q} \rangle$ and $d_{\mathbf{k}}$. {In what follows, we fix $t^f=-0.3$ and consider a half-filled band case, i.e., }
\begin{equation}
\label{3}
n=\langle n^f\rangle+\langle n^c\rangle =1\,,
\end{equation}
where 
\begin{equation}
\langle n^f\rangle=\frac{1}{N}\sum_{\bf k} \langle f^\dagger_{\bf k}f^{}_{\bf k}\rangle,  
\end{equation}
and 
\begin{equation}
\langle n^c\rangle
=\frac{1}{N}\sum_{\bf k} \langle c^\dagger_{\bf k}c^{}_{\bf k}\rangle,
\end{equation}
are respectively the densities of valance and conduction electrons. The chemical potential $\mu$ therefore has to be adjusted in such a way that Eq.~(\ref{3}) is satisfied. The difference between the on-site energies of a $c$-electron and an $f$-electron is $\varepsilon^c-\varepsilon^f=1$. 

\begin{figure}[htb]
\begin{center}
     \includegraphics[angle = 0, width = 0.45\textwidth]{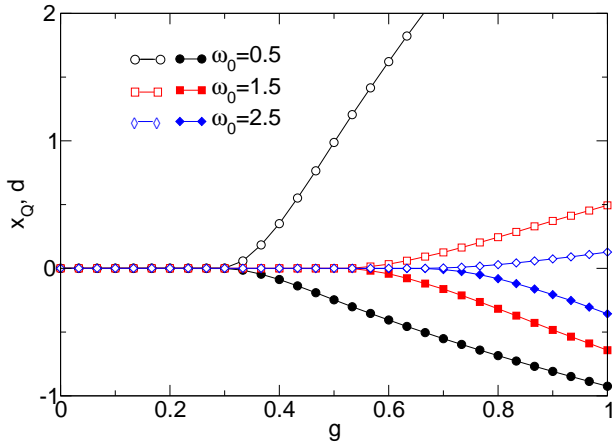}
\end{center}
\caption{EI order parameter $d$ (filled symbols) and lattice displacement $x_{\mathbf{Q}}$ (open symbols) as functions of electron-phonon coupling $g$ for different phonon frequencies $\omega_0$ at zero temperature.}
\label{fig:1}
\end{figure}

{As a first step}, we discuss {the ground state} of the electron-phonon systems modeled by the Hamiltonian in Eq.~(\ref{1}). Figure~\ref{fig:1} shows us a dependence of the EI order parameter $d$ and the lattice displacement $x_{\mathbf{Q}}$ on the electron-phonon coupling $g$ at zero temperature. We see that $d$ and $x_Q$ are intimately related. For a given value of phonon frequency $\omega_0$, $d$ and $x_Q$ appear to be nonzero only if the exciton-phonon coupling is larger than a critical value $g_c$. {This $g_c$ increases when} increasing the phonon frequency. This feature has been addressed in the one-dimensional case in which a linear relation between a square critical coupling $g_c$ and the phonon frequency has been confirmed both in numerical and analytical calculations~\cite{PBF13}. {In the case of} $g>g_c$, the EI/CDW state exhibits a finite lattice distortion. {In the contrast, i.e., with $g<g_c$,} the system settles in a semi-metal state with an undistorted lattice structure. This semimetal EI/CDW transition is Kosterlitz-Thouless type, typically observed in general two-dimensional systems~\cite{KT73}. Note here that in {the} one-dimensional case, the model written in Eq.~(\ref{1}) also consists the Kosterlitz-Thouless transition form of the semimetal EI {transition~\cite{PBF13}.}

\begin{figure}[htb]
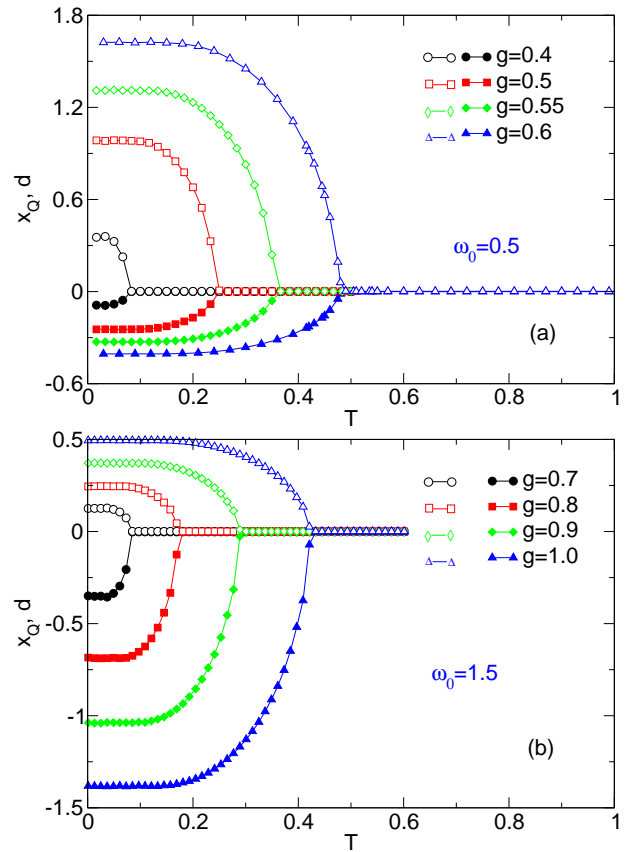

\begin{center}
     \includegraphics[angle = 0, width = 0.45\textwidth]{fig2a.eps}
     \includegraphics[angle = 0, width = 0.45\textwidth]{fig2b.eps}     
    \end{center}
\caption{EI order parameter $d$ (filled symbols) and lattice displacement $x_{\mathbf{Q}}$ (open symbols) as functions temperature $T$ for different electron-phonon couplings $g$ at phonon frequency $\omega_0=0.5$ (a) and $\omega_0=1.5$ (b).}
\label{fig:2}
\end{figure}

Figure~\ref{fig:1} indicates that our situation can be split into two different regimes characterized by the phonon frequency $\omega_0$. In the adiabatic regime ($\omega_0<t$), for instance $\omega_0=0.5$, the lattice displacement is larger than the EI order parameter, whereas in the opposite case, i.e., in the anti-adiabatic regime ($\omega_0>t$) the lattice displacement is smaller and goes {more slowly} than the EI order parameter. {For that reason, the} temperature effect on the EI/CDW semimetal transition needs to be discussed separately in both regions. In the following we focus on two typical values of $\omega_0$, $\omega_0=0.5$ and $\omega_0=1.5$.
 
In Figure~\ref{fig:2}(a) the EI order parameter $d$ and the lattice displacement $x_\mathbf{Q}$ are shown as functions of temperature for some values of $g$ at $\omega_0=0.5$. Obviously, the EI state stabilizes at low temperature. That is analogous to the case of superconductivity. The EI order parameter $d$ decreases if temperature is {increased}. It completely disappears at a critical temperature, $T_c$. $T_c$ here therefore stands for an EI/CDW transition temperature. The EI transition temperature is weakened by lowering the coupling of the electron-phonon interaction. This once more {reminds us of} the BCS theory in which the condensation of Cooper pairs is driven by the electron-phonon interaction. The BCS type condensation of electron-hole pairs in the present problem will be discussed in detail in the remained figures. The temperature dependence of $d$ shown in Figure~\ref{fig:2}(a) here fits quite well with the recent experimental observation in the quasi-two dimensional $1T$-TiSe$_2$ system~\cite{MSGMDBACMBBT10}. That good description of the mean-field approach strongly indicates the macroscopic condensation of coherent excitons below the critical temperature. Above the critical temperature, the strong fluctuations of incoherent electron-hole pairs would be significant~\cite{MSGMDBACMBBT10}
\begin{figure}[htb]
    \begin{center}
     \includegraphics[angle = 0, width = 0.45\textwidth]{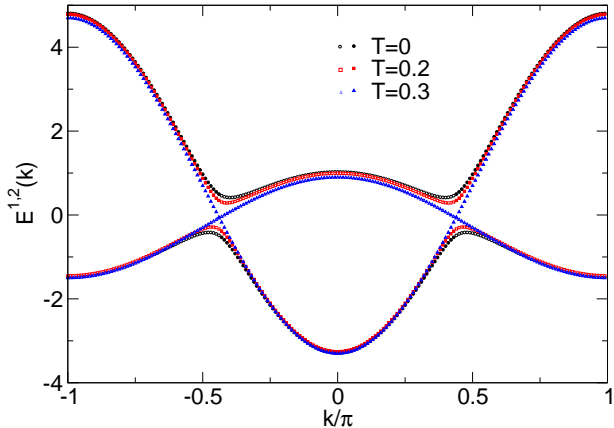}
    \end{center}
\caption{Quasiparticle band dispersions for $c$-electrons, $E^1(\mathbf{k})$ and $f$-electrons, $E^2(\mathbf{k})$ along $(k,k)$ direction for different temperatures $T$ at $g=0.5$ and $\omega_0=0.5$.}
\label{fig:3}
\end{figure}

\begin{figure*}[htb]
    \begin{center}
     \includegraphics[angle = 0, width = 0.29\textwidth]{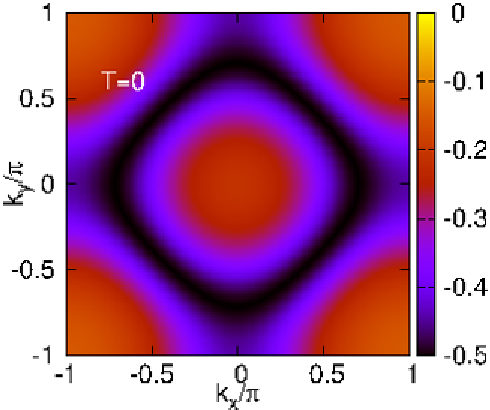}
     \includegraphics[angle = 0, width = 0.29\textwidth]{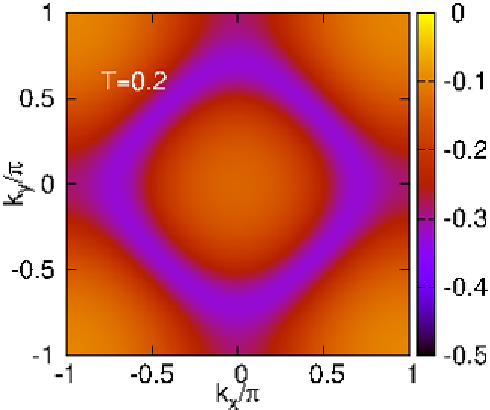}
     \includegraphics[angle = 0, width = 0.31\textwidth]{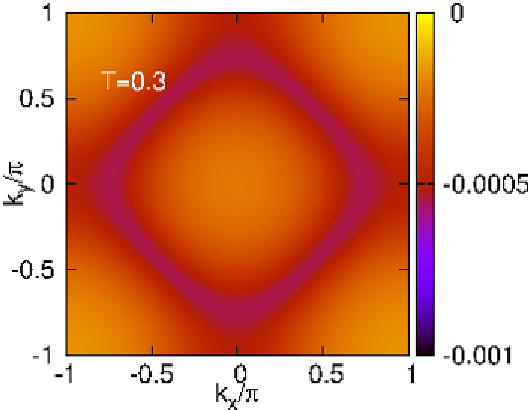}
\end{center}
\caption{Magnitude of the EI order parameter $d_{\mathbf{k}}$ depending on momentum $\mathbf{k}$ in the whole first Brillouin zone for different temperatures at $g=0.5$ and $\omega_0=0.5$.}
\label{fig:4}
\end{figure*}

In Figure~\ref{fig:2}(a) we show also the lattice displacement $x_\mathbf{Q}$ evaluated for the same set of parameters. As in the ground state (cf. Figure~\ref{fig:1}), at finite temperature we still recognize the analogous behaviour of the lattice displacement and {the} EI order parameter. They simultaneously disappear if {the} temperature is larger than the EI/CDW transition temperature. Below the transition temperature, the lattice distortion occurs according to the stability of the exciton condensation. This temperature dependence of the displacement in the effect of the electron-phonon interaction has been discussed in the tight-binding approximation where the mean-field like temperature dependence of {the} EI order parameter has been assumed~\cite{MBCAB11}. In our task both lattice displacement and EI order parameter are calculated self-consistently, i.e., no manual form of them needs to be chosen at the beginning. The temperature dependence of the lattice displacement agrees {qualitatively} with extracted data from neutron diffraction experiments at low temperatures below $T_c$~\cite{SMW76}. That once more confirms the validity of the mean-field approach at low temperature to study the lattice distortion induced by the EI/CDW state. {Similarly to Figure~\ref{fig:2}(a)}, Figure~\ref{fig:2}(b) displays the temperature dependence of the EI order parameter and the lattice displacement for some $g$ values but in the anti-adiabatic case. In this situation the lattice displacement is always smaller than the EI order parameter in the whole range of temperature. In what below, we focus on the adiabatic regime to discuss {the} effects of temperature on the microscopic properties in our system. In the {anti-adiabatic case} we find (not shown here) the same physical scenarios.

Figure~\ref{fig:3} shows the renormalized quasiparticle energy bands $E_{\mathbf k}^1$ and $E_{\mathbf k}^2$ along the diagonal direction of the first 2D Brillouin zone, i.e.,~$k_x=k_y$, for  different temperatures at $g=0.5$ and $\omega_0=0.5$. Note that we have chosen the semi-metallic situation meaning that in the non-interacting case, both {$c$- and $f$-bands overlap. The} Fermi surface in this case is large and both types of quasiparticles participate to form the Fermi surface. {At low temperature, it shows us that a sufficiently large electron-phonon coupling triggers a gap opening at the Fermi level}. {The opening of the gap indicates the bound state of the electron-hole pairs. This feature makes manifest the experimental observation in TmSe$_{0.45}$Te$_{0.55}$ that at large pressure, two $4f$- and $5d$-bands overlap. Due to the phonon scattering, $4f$-holes couple to $5d$-electrons to form excitons. At sufficiently low temperature, these excitons condense~\cite{Wa01}}. This once again {reminds of} a similar relevance to the BCS theory of the superconductivity where Cooper pairs are formed. {The} width of the gap is proportional to the EI order parameter, which therefore decreases when enhancing temperature {(cf. Fig.~\ref{fig:2}). Below the EI/CDW transition temperature electron-hole pairs are formed. Whereas, above the EI/CDW transition temperature, large thermal fluctuations destroy the bound electron-hole pairs and the gap disappears}. For a strictly 2D system, of course, the critical temperature for exciton condensation would be zero, but the superfluid properties should survive for temperatures smaller than the Kosterlitz-Thouless transition temperature~\cite{ZJ08,MBSM08,PF12}.

Next, in Figure~\ref{fig:4} we address {the} momentum dependence of the order parameter $d_\mathbf{k}=\langle c_{\mathbf{k}+\mathbf{Q}}^\dag f_\mathbf{k}\rangle$ in the first Brillouin zone for different temperatures at the same set of parameters given in Figure~\ref{fig:3}. Note that besides $d$, $d_\mathbf{k}$ also characterizes  the EI order parameter. For low temperatures and ${\bf k}$ close to the Fermi momentum, $d_{\mathbf k}$ is strongly peaked, otherwise $d_{\mathbf k}$ is a rather smooth function of $\mathbf k$. As a matter of course, increasing temperature, {the} amplitude of the peaks {goes down} and it becomes insignificant if {the} temperature $T$ {is higher} than the EI transition temperature. At low temperature, Fermi surface apparently plays an important role to form the electron-hole bound state, indicating the typical BCS-type of the EI stability~\cite{PBF10,PFB11}.

\begin{figure}[t]
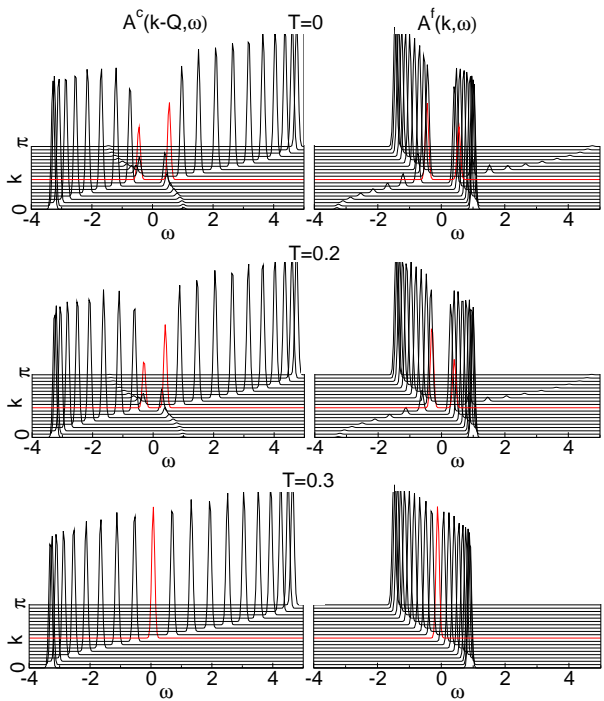

    \begin{center}
     \includegraphics[angle = 0, width = 0.45\textwidth]{fig5a.eps}\\
     \includegraphics[angle = 0, width = 0.45\textwidth]{fig5b.eps}\\
     \includegraphics[angle = 0, width = 0.45\textwidth]{fig5c.eps}
    \end{center}
\caption{Single particle spectral functions of $c$-electron (left) and $f$-electron (right) along $(k,k)$ direction. The set of parameters are given in Figure~\ref{fig:3}. Red lines indicate the spectral function at Fermi momentum.}
\label{fig:5}
\end{figure}

To understand {more} about the temperature effects in the electron-hole bound state we now present the wave-vector and frequency resolved single-particle spectral functions associated with the photoemission or inverse photoemission (injection) of $c$ and $f$ electrons. Figure~\ref{fig:5} shows the variation of $A^{c}(\mathbf{k}-\mathbf{Q},\omega)$ (left) and $A^f(\mathbf{k},\omega)$ (right) for different temperatures $T$ as discussed in the last two figures following the high-symmetry diagonal direction in the Brillouin zone. 

At zero temperature (see top panels) we find the gap feature opened at {the} Fermi level (see also Figure~\ref{fig:3}). In this case, $c$- and $f$-electron states strongly hybridize close to the Fermi energy with large spectral weight transferring. {Electron-hole pairs are therefore} created and then condense {in BCS type}. The single-particle excitation gap is suppressed by increasing temperature (see middle panels) and completely disappears when {the} temperature is larger than the critical value (see bottom panels). The gapless excitation at high temperature indicates that the bounding state of electron-hole pairs is broken. At high temperature, the system settles in the electron hole plasma state. 

\begin{figure}[t]
    \begin{center}
     \includegraphics[angle = 0, width = 0.23\textwidth]{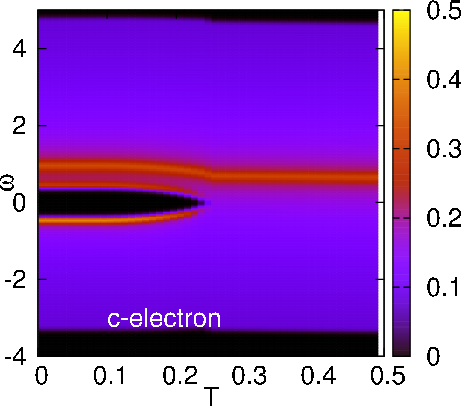}
     \includegraphics[angle = 0, width = 0.23\textwidth]{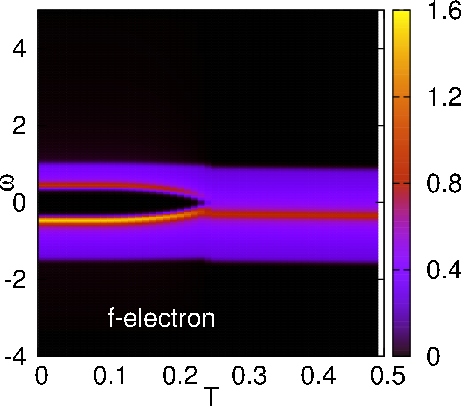}
    \end{center}
\caption{DOS intensity of $c$-band (left) and $f$-band (right) as functions of temperature at $g=0.5$ and $\omega_0=0.5$.}
\label{fig:6}
\end{figure}

To analyze the dynamical properties of the electron-hole systems in the full range of temperature we discuss below a scenario of density of state (DOS). In Figure~\ref{fig:6} intensity plots are shown for {the} DOS of the $c$-band (left panel) and $f$-band (right panel) as functions of temperature at $g=0.5$ and $\omega_0=0.5$. At high temperature the systems clearly {show metallic state} with nonzero DOS at the Fermi energy. Lowering temperature, a correlation induced ``hybridization'' gap opens, indicating long-range order of {the} non-vanishing $f-c$ polarization and also the lattice displacement. {A strong enhancement} of the DOS at the upper valence and lower conduction band edges due to the $c-f$-mixing state {reminds us of} a BCS-type structure evolving from a semimetallic state with a large Fermi surface above $T_c$~\cite{IPBBF08}. 

The microscopic investigations of the system depending on temperature are in qualitatively good agreement with experimental observations~\cite{CMCBDGBA07,MSGMDBACMBBT10}. Our results indicate that the excitons in the system are condensed in a BCS-like manner and give rise to a CDW state, characterized by the signatures of the EI order parameter and the lattice displacement as discussed before.

\section{Summary}
To summarize, in this paper we have developed the unrestricted Hartree-Fock approximation adapting to the two-band electronic-phonon interaction model to address {the} influence of temperature on the excitonic condensate and lattice displacement in quasi two-dimensional systems. With an expectation that the electron-hole pair bound state might be formed at low temperature we have derived self-consistent equations permitting us to determine the EI order parameter and the lattice displacement. Numerical results show us that, in the ground state, the EI stability and the lattice displacement are intimately related at large electron-phonon interaction. Below a critical value of the electron-phonon coupling strength the system settles in the metallic state. That happens for both adiabatic and anti-adiabatic limitations. Extending the calculation over a wide temperature range, we find a critical temperature at which the bound state of electron-hole pairs start to be completely destroyed by thermal fluctuations. By inspecting in further detail the momentum dependence of quasi-particle bands, the EI order parameter and photoemission spectral functions, we once more confirm the BCS type condensation of the electron-hole pairs at low temperature. {In particular}, the photoemission spectral function reveals a pronounced back-folding of the spectral signature in the EI state that directly makes manifest the transition forward a CDW state. Analyzing the temperature dependence of $c$ and $f$ electron DOSs also exhibits the striking BCS type structure of the EI state evolving from a semimetallic state with a large Fermi surface above $T_c$. In this way our work has pointed out the prominent role played by the lattice degrees of freedom establishing a charge-density wave in semimetallic systems with weak (indirect) band overlap and in mixed-valent semi-conductors with band gaps comparable to the exciton binding energy, such as quasi-two-dimensional 1$T$-TiSe$_2$. The good agreement with experimental observations of our results strongly indicates that the unrestricted Hartree-Fock approach is applicable to consider excitonic condensation inducing the charge density wave state with  lattice distortion. Our work therefore gives strong support for exciton condensation as a purely phonontic mechanism responsible for the CDW phase in materials. Combining with the electronic mechanism (due to Coulomb interactions) to consider competition of the EI/CDW state and the lattice distortion will be left to the future.
\acknowledgements
The authors would like to thank Prof. H. Fehske for valuable discussions. This research is funded by the Vietnam National Foundation for Science and Technology Development (NAFOSTED) under grant number 103.01-2014.05.

\bibliographystyle{apsrev}
\end{document}